\def\lesssim{\mathrel{\hbox{\rlap{\hbox{\lower4pt\hbox{$\sim$}}}\hbox{$<$}}}}
\def\gtrsim{\mathrel{\hbox{\rlap{\hbox{\lower4pt\hbox{$\sim$}}}\hbox{$>$}}}}
\def\msun{$M_{\odot}$}
\def\teff{$T_{\rm eff}$~}
\def\lteff{log${T_{\rm eff}}$~}
\def\ll_lsun{log$({L/\rm L_{\sun}})$~}
\def\masa_msun{$M/ \rm M_{\sun}$~}
\def\m_mstar{$M/M_{*}$~}

\def\lteff{log\ $T_{\rm eff}$}
\def\teff{$T_{\rm eff}$}

\def\gteff{$\log$ \teff $-$$\log g$}

\documentclass{aa}

\usepackage{graphicx}
        
\begin{document}

\title{On the robustness of H-deficient post-AGB tracks}

\author{M. M. Miller Bertolami$^{1,2}$\thanks{Fellow of CONICET, Argentina.},
        L. G. Althaus$^{1,2}$\thanks{Member of the Carrera del Investigador
        Cient\'{\i}fico y Tecnol\'ogico, CONICET, Argentina.}}
\offprints{M. M. Miller Bertolami }

\institute{Facultad de Ciencias  Astron\'omicas y Geof\'{\i}sicas,
           Universidad  Nacional de  La Plata,  
           Paseo del  Bosque s/n,
           (1900) La Plata, Argentina.\
           \and
           Instituto de Astrof\'{\i}sica La Plata, IALP, CONICET\\
\email{mmiller,althaus@fcaglp.unlp.edu.ar}}
\date{\today}

\abstract{}
         {We analyze the robustness of H--deficient post--AGB tracks
regarding previous evolution of their progenitor stars and the
constitutive physics of the remnants. Our motivation is a recent
suggestion of Werner \& Herwig (2006) that previous evolution should
be important in shaping the final post--AGB track and the persisting
discrepancy between asteroseismological and spectroscopical mass
determinations. This work is thus complementary to our previous work
(Miller Bertolami \& Althaus 2006) and intends to shed some light on
the uncertainty behind the evolutionary tracks presented there.}  {We
compute full evolutionary models for PG1159 stars taking into account
different extramixing (overshooting) efficiencies and lifetimes on the
TP-AGB during the progenitor evolution. We also assess the effect of
possible differences in the opacities and equation of state by
artificially changing them before the PG1159 stage. Also comparisons
are made with the few H-deficient post--AGB tracks available in the
literature.}  {Contrary to our expectations, we found that previous
evolution is not a main factor in shaping H--deficient post--AGB
tracks. Interestingly enough, we find that only an increase of
$\sim50\%$ in the intershell opacities at high effective temperatures
may affect the tracks as to reconcile spectroscopic and
asteroseismologic mass determinations.  This forces us to conclude
that our previous tracks (Miller Bertolami \& Althaus 2006) are robust
enough as to be used for spectroscopic mass determinations, unless
opacities in the intershell region are substantially different. Our
results, then, call for an analysis of possible systematics in the
usually adopted asteroseismological mass determination methods.}  {}
\authorrunning{Miller Bertolami \& Althaus}
\titlerunning{On the robustness of H--deficient post--AGB tracks}

\maketitle


\section{Introduction}
Post Asymptotic Giant Branch (AGB) stars constitute a short--lived
transition stage between AGB stars and white dwarf stars. Among them a
minority show H--deficient compositions and are suppossed to be the
main progenitors of H--deficient white dwarfs, which account for about
15\% of the white dwarf population (Eisenstein et al. 2006). The group
of H--deficient post--AGB stars displays a wide variety of surface
chemical compositions ranging from almost pure helium envelopes to the
helium-- (He), carbon-- (C) and oxygen-- (O) rich surface composition
of the Wolf Rayet central stars of planetary nebulae ([WC]) and the
PG1159 type stars; see Werner \& Herwig (2006), from now on WH06.
 The surface composition of the last group resembles the
intershell region chemistry of AGB star models when some overshooting
at the base of the pulse driven convective zone (PDCZ) is allowed
during the thermal pulses (Herwig et al. 1997). For this reason, and
also due to the fact that the occurrence of late (i.e. post-AGB)
thermal pulses is statistically unavoidable in single stellar
evolution modeling (Iben et al. 1983), a late helium shell flash is
the most accepted mechanism for the formation of these stars (see,
however, De Marco 2002). In this scenario, 
the remaining thin H--rich envelope is either burnt in a very
late helium flash (VLTP) that occurs on the hot white dwarf cooling
branch after H burning has almost ceased, or diluted in a late helium
flash (LTP) that develops when the H burning shell is still active
during the horizontal evolution of the stars in the HR diagram (Herwig
2001).

Roughly a third of PG1159--type stars exhibit multiperiodic luminosity
variations caused by non--radial g--mode pulsations. This has allowed
researchers to derive structural parameters --- in particular the mass 
of these stars --- of individual pulsators by means of asteroseismological
studies i.e. by comparing adiabatic pulsation periods with the observed
ones --- e.g. Kawaler \& Bradley (1994) and more recently C\'orsico \&
Althaus (2006). It is important to mention that for applications
requiring accurate values of adiabatic pulsation periods 
full evolutionary models with a realistic thermal structure should be
used. Stellar masses of PG1159 stars can also be derived by comparing
the values of log $ g $ and \lteff\ coming from the fitting of
line--blanketed non--LTE model atmospheres to the measured spectra
(Werner et al. 1991) with tracks coming from stellar evolution
modeling. These two different approaches enable us to compare the
derived stellar masses. Although previous spectroscopical mass
determinations, based on old H--rich post--AGB models, show relatively
good agreement with asteroseismological masses (to about 5\%, WH06, roughly 
0.03 \msun), the development of a new generation
of stellar evolution sequences that account for the C-- and O-- rich
surface abundances expected in PG1159 stars (Herwig et al. 1999) has
changed the situation. As mentioned by WH06 the new post--AGB tracks
are systematically hotter than the old ones, which leads to lower
spectroscopical masses. The new mean spectroscopical mass becomes
0.573 \msun, this is 0.044 \msun\ lower than previous values; see
Miller Bertolami \& Althaus (2006), from now on MA06. This is at variance with
asteroseismological predictions.  In fact from Table 3 of WH06 and Table
2 of MA06 the asteroseismological masses are usually 10\% higher than their 
spectroscopical counterparts, except for the hottest known  pulsating
PG1159 star \object{RX J2117.1+3412}, the spectroscopical mass of which is more than 
20\% higher than the asteroseismological one; see 
C\'orsico et al. (2007) for a recent and detailed study of this object. 
The difference in derived masses is a clear indication
of the uncertainties weighting upon the mass determination methods. 
%
%

In this context, WH06 have recently compared new and old tracks and
claimed that the previous evolution on the thermally pulsing AGB
(TP-AGB) --- particularly the third dredge-up (3DUP)
efficiency
and mass--loss rates --- plays a decisive role in the location of the
tracks in the HR and \gteff\ diagrams during the post--AGB
evolution. Specifically, as shown by Herwig et al. (1998), a strong
3DUP changes the evolution of the core mass without altering the
evolution of its radius. Consequently the mass--radius relation of the
remnants will depend on the previous TP-AGB evolution and, if we
accept in the ``prediction'' of shell homology relations ($ L_{\rm
shell} \sim M_{\rm core}^2 {R_{\rm core}}^{-1}$, for $M_{\rm core}
\lesssim 0.8$\msun, Herwig et al. 1998), the post--AGB tracks would be
accordingly altered. WH06 also point out that mass loss can produce a
similar effect as remnants of similar mass may come out with very
different degrees of degeneracy depending on the previous
evolution. This being the case, as both mass--loss rates and 3DUP
efficiency are poorly known, the location of theoretical post--AGB
tracks, and thus mass determination, would be highly model dependent
and uncertain.  These issues call for the need of an analysis of the
robustness of existing H-deficient post--AGB tracks and for a way of
solving the mentioned mass discrepancies

 However, no calculation of the importance of these effects was
actually presented neither in WH06 nor in Herwig et al. (2006). The
lack of consistent calculations to assess to what a degree the
location of the post--AGB tracks depends on the prior AGB evolution
has motivated us to undertake the present investigation.  In the
following sections we will elaborate on these issues. In this sense
the present work is complementary to that of MA06 where H-deficient
post--AGB tracks were presented but no analysis of its robustness was
performed. In Sect. 2 we analyze how evolution previous to the PG1159
stage affects PG1159 tracks in light of the suggestion presented by
WH06. Then in Sect. 3 we explore to what an extent the constitutive
physics of the models at the PG1159 stage may affect the tracks.  In
Sect. 4 we compare with other H-deficient tracks available in the
literature and also compare the location of tracks coming from LTP and
VLTP events. Finally Sect. 5 is devoted to the discussion of the
results and making some final remarks.

\section{Influence of previous evolution}
As was mentioned, uncertainties in mass--loss rates are expected to
affect the duration of the TP-AGB phase and to lead to remnants with
different degrees of degeneracy and mass--luminosity relation. Also
the initial-final mass relation is expected to be altered by different
mass--loss rates. By altering the intensity of mass loss we can get
the same final remnant mass from progenitors of initially very
different mass, and consequently very different previous evolutions
(e.g., that have or have not undergone a helium core flash at the tip
of the RGB). We will address these issues in Sect. 2.1. In Sect. 2.2
we elaborate on the effects of different 3DUP efficiencies on the
TP-AGB, which is the other point mentioned in WH06 as a possible cause
for shifts in post--AGB tracks. The main effect of 3DUP efficiency is
to change the initial-final mass relation. Indeed, strong dredge up
events on the TP-AGB lead to lower final remnant masses for the same
initial mass. In this context we analyze sequences with different 3DUP
efficiencies and, to disentangle this effect from the one studied in
Sect 2.1, with the same TP-AGB lifetime. Finally, Sect 2.3 is intended
to clarify the reason of the difference between MA06 and Bl\"ocker
(1995a,b) tracks and to study the extreme  limiting case for
which no overshooting (OV) is allowed to operate at any convective
boundary during the whole evolution.  In all the sequences
presented in this section, mass loss has been arbitrarily set during the
departure from the AGB as to get a VLTP and the subsequent
PG1159-like surface composition.

To visualize and quantify the change introduced by the variations
in the parameters of each sequence we will refer and compare our
sequences with those of MA06 and C\'orsico et al. (2006) which are
assumed as standard in the present work. These sequences were
calculated with an overshooting efficiency of $f=0.016$ at all
convective boundaries; see Herwig et al. (1997) for a definition of
$f$. To quantify the change that a variation in \teff\ and $g$ for a
sequence of a given mass --- caused by different physical assumptions
in the calculations --- would produce in spectroscopical mass
determinations, we estimate a mass value for the sequence from its
location relative to MA06 sequences --- this is what is called ``mass
derived from comparison'' in Table 1 --- and compare that mass with
the actual value of the mass. The difference between both values gives
the shift expected in spectroscopical masses if tracks with a
different physical assumption are used in their derivations.

  It is worth noting that most of our article deals with
post-VLTP sequences. However, within the late helium flash scenario
for explaining the origin of PG1159 stars, these objects could also be
the offspring of LTP events. In fact some PG1159 stars are known to be
$^{14}$N-deficient, a fact usually asociated with post-LTP objects. In
these cases some H will be present but hidden below the detection
limit. If systematic differences exist between post-LTP and post-VLTP
tracks, then this will introduce a systematic effect in spectroscopic
mass determiations. Although from figure 1 of Herwig one is tempted to
conclude that this is not the case, it is worth noting that the presence of H
should be more important in the low mass region as these stars display
thicker H-envelopes. We will discuss this issue in section 4.3, where
detailed comparisons between post-LTP and post-VLTP tracks will be
made for a wide range of masses and various surface H-abundances.
.

\begin{table}[ht!]
\begin{center}
\begin{tabular}{|c|c|c|c|}\hline
Sequence & Final Mass & Mass derived \\ & & from comparison \\\hline
 NALT & 0.607 & 0.612 \\ LALT & 0.6035 & 0.614 \\ SALT & 0.6033 &
 0.598 \\ 2.2MSALT & 0.5157 & 0.524 \\ TPA008 & 0.617 & 0.621 \\
 TPA004 & 0.633 & 0.637 \\

3\msun\ w/NOV & 0.626 & 0.623 \\ \hline
\end{tabular}
\label{tab:masses}
\caption{Values of the final masses of the sequences of this work
and the masses derived from the comparison with the
``standard'' ones (MA06). Stellar masses are in \msun. See Sect. 2.1, 2.2, and 2.3 for definition of 
the sequences.}
\end{center}
\end{table}
\subsection{Effect of different TP-AGB lifetime or mass--loss rate}
As stated in WH06, for a similar core mass, a
remnant that spend more time on the TP-AGB will finish with a more
compact and degenerate core. Then, different mass--loss prescriptions can lead
to remnants with the same core mass but different radius and, by virtue of
shell homology relations --- that ``predict'' $L_{\rm shell} 
\sim M_{\rm core}^2 R_{\rm core}^{-1}$, Herwig et al. (1998) --- 
different luminosities. This is supported by the work of Herwig et
al. (1998) that shows that, because of the continuous shrinking of the
H-free core (HFC), the luminosity at the TP-AGB keeps increasing,
despite the end of the effective core mass growth as consequence of
strong dredge up events. In addition, Bl\"ocker (1995b) has already
shown that a more compact remnant is more luminous than a less compact
one of similar mass.  To analyze the possible shift in the H-deficient
post-AGB tracks resulting from uncertainties in TP-AGB mass loss ---
and hence in different TP-AGB lifetimes --- we have calculated the
full evolution of three sequences with the same prescriptions as in
MA06 but changing mass loss at the TP-AGB to get different TP-AGB
lifetimes. These sequences are: NALT with a normal mass--loss
prescription, SALT with a short TP-AGB lifetime and LALT with a longer
TP-AGB. All these sequences come from the same pre TP-AGB evolution of
an initially 3-\msun\ ZAMS star. While NALT underwent 12 thermal
pulses, SALT and LALT sequences experienced 6 and 18 pulses,
respectively.  SALT (LALT) sequence has a TP-AGB lifetime a factor 2
shorter (1.5 longer) than NALT. Thus the sequences considered here
take into account possible uncertainties in TP-AGB lifetimes up to a
factor three. This is more than what is expected from different
mass--loss prescriptions (Kitsikis \& Weiss 2007). Due to the high
dredge up efficiency during the last thermal pulses the final remnant
mass of all these sequences is very similar (see Table 1), thus
allowing a direct comparison of the effect of different TP-AGB
lifetimes on the location of H-deficient post--AGB tracks of similar
mass.  We mention that all of these sequences have been followed
through an additional post-AGB thermal pulse (the VLTP) where the
H-rich envelope is violently burned.

\begin{figure}
  \includegraphics[clip, width=250 pt]{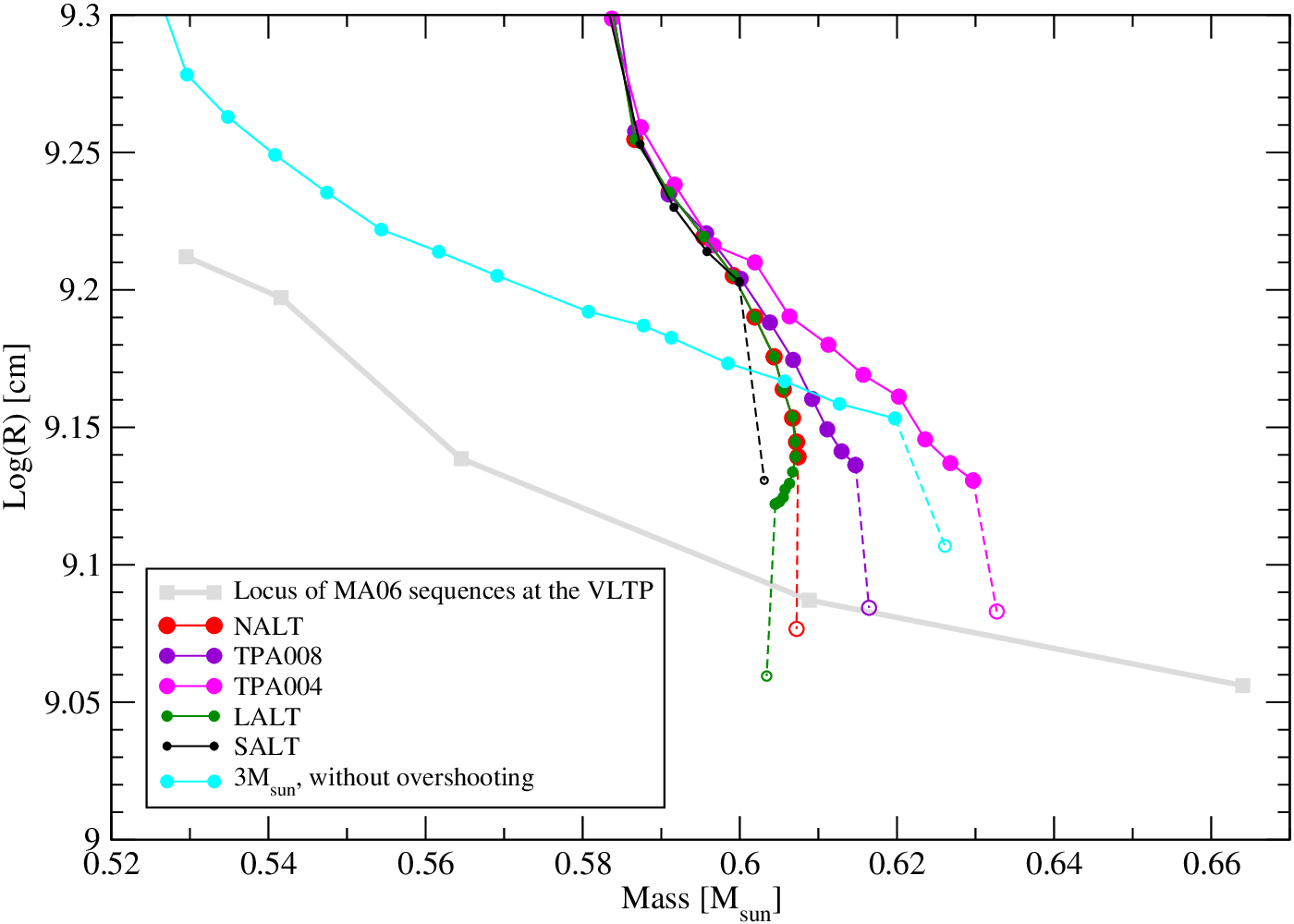} \caption{Evolution of
  the HFC (mass and radius) during the TP-AGB (solid lines,
  filled circles) and at the VLTP (dashed lines, empty circles) for
  selected sequences (values are taken just before
  each thermal pulse). Also the locus of the standard models at the moment just
  before the VLTP is shown for comparison. Note that, due to the turn off of
  the H-burning shell, compression before the VLTP does not follow the 
  trend in the AGB. [color figure only available in the electronic version]}
\label{fig:Hfreecore} 
\end{figure}

\begin{figure}[ht!]
  \includegraphics[clip, width=250 pt]{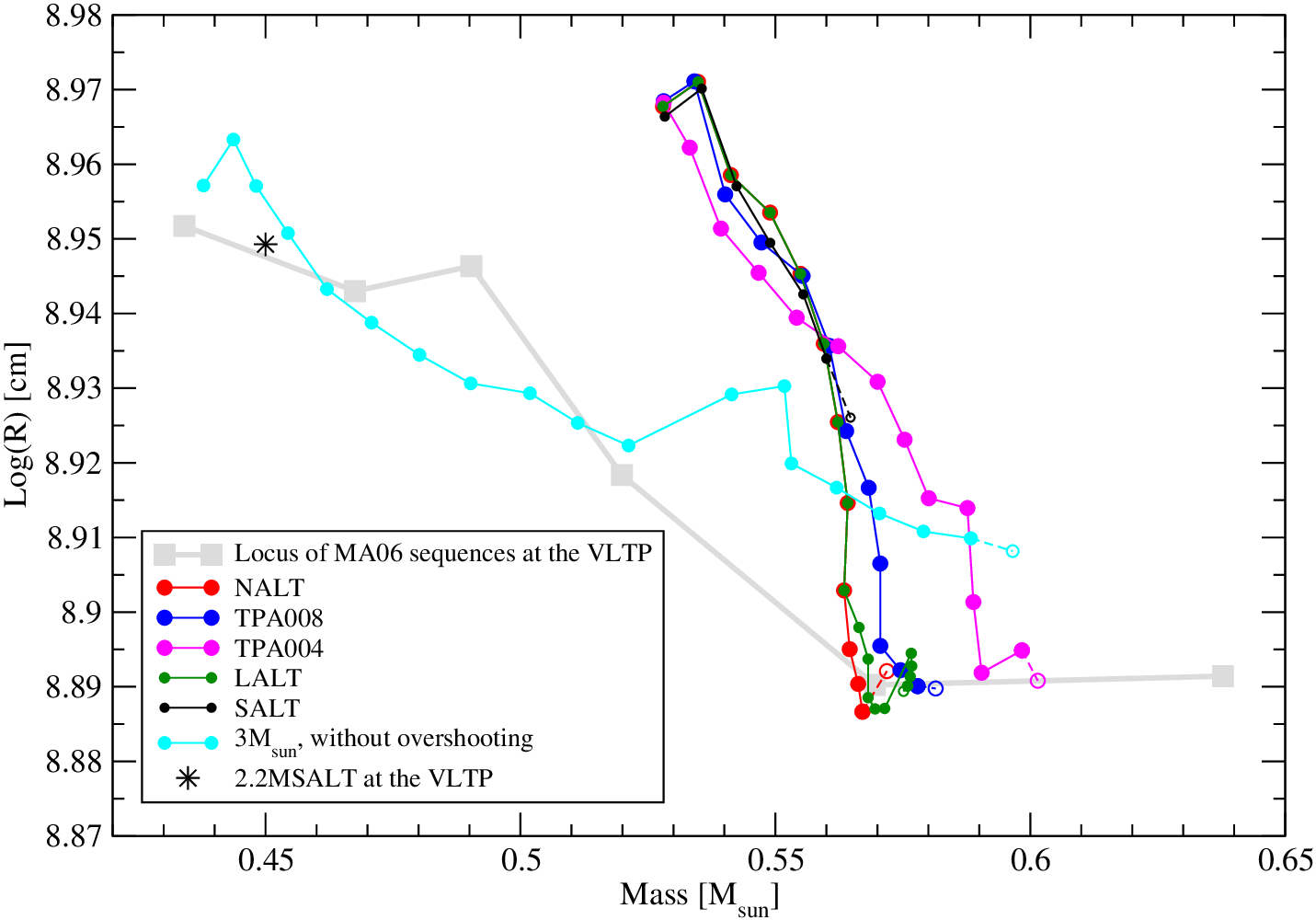} 
\caption{Same as Fig. \ref{fig:Hfreecore} but for the HeFC.
Note that the location of the HeFC on this
diagram seems to converge, after not many thermal pulses, to a certain 
locus. [color figure only available in the electronic version]}
\label{fig:Hefreecore} 
\end{figure}

In some agreement with Bl\"ocker (1995b) we find a shift in post--AGB
tracks as a consequence of different TP-AGB lifetimes. However the
effect is not very important. In fact, comparing SALT and LALT sequences (both
with the same final mass) we see that a factor 3 in TP-AGB lifetimes
leads to a maximum shift of 0.03 dex in \lteff. A shorter TP-AGB leads
to cooler tracks that would imply $\sim0.015$ \msun\  larger 
spectroscopical masses. It is also worth noting that tracks for LALT
and NALT sequences are almost identical regardless the difference in
TP--AGB lifetime of 50\%. It seems that, while shortening the TP--AGB
does change the post--VLTP tracks, prolonging it does not produce a
sizeable effect. To understand this, we show in Figs. \ref{fig:Hfreecore} 
and \ref{fig:Hefreecore} the mass-radius relations of our sequence for both 
the H-- and He-- free cores --- HFC and HeFC, respectively. The evolution of 
the HFC is in agreement with that
presented by Herwig et al. (1998) which shows that the HFC continues to
contract even when the core mass growth is stopped by efficient 3DUP
events. Because this behaviour is the basis of the argument of WH06 
the following should be noted. First the radius of the HFC at the 
moment of the
VLTP does not follow the trend during the TP-AGB. This is a result of
the accelerated compression of the intershell caused by the decline 
of the H-burning shell when
the star approaches the white dwarf cooling track. Second, and more
importantly, the post--VLTP sequences are powered by the
He-burning shell and consequently, if shell homology are to be used in the
analyzes, the relevant values should be the \emph{HeFC mass and
  radius}. Note that the HeFC (Fig. \ref{fig:Hefreecore}) seems to converge to
a certain locus in the core mass-radius diagram faster than the HFC. In fact
while in all the sequences the HFC radius gets smaller with each thermal
pulse, the HeFC ends its compression after about $\sim 10$ thermal pulses.
This helps to understand why there is almost no difference between NALT
and LALT sequences. The 6 ``extra'' thermal pulses of LALT sequence 
do not introduce any
significant change in the mass-radius relation and thus, according to
shell homology relations, their  post-AGB luminosity should be similar.

\begin{figure}[t]
  \includegraphics[clip, width=260 pt]{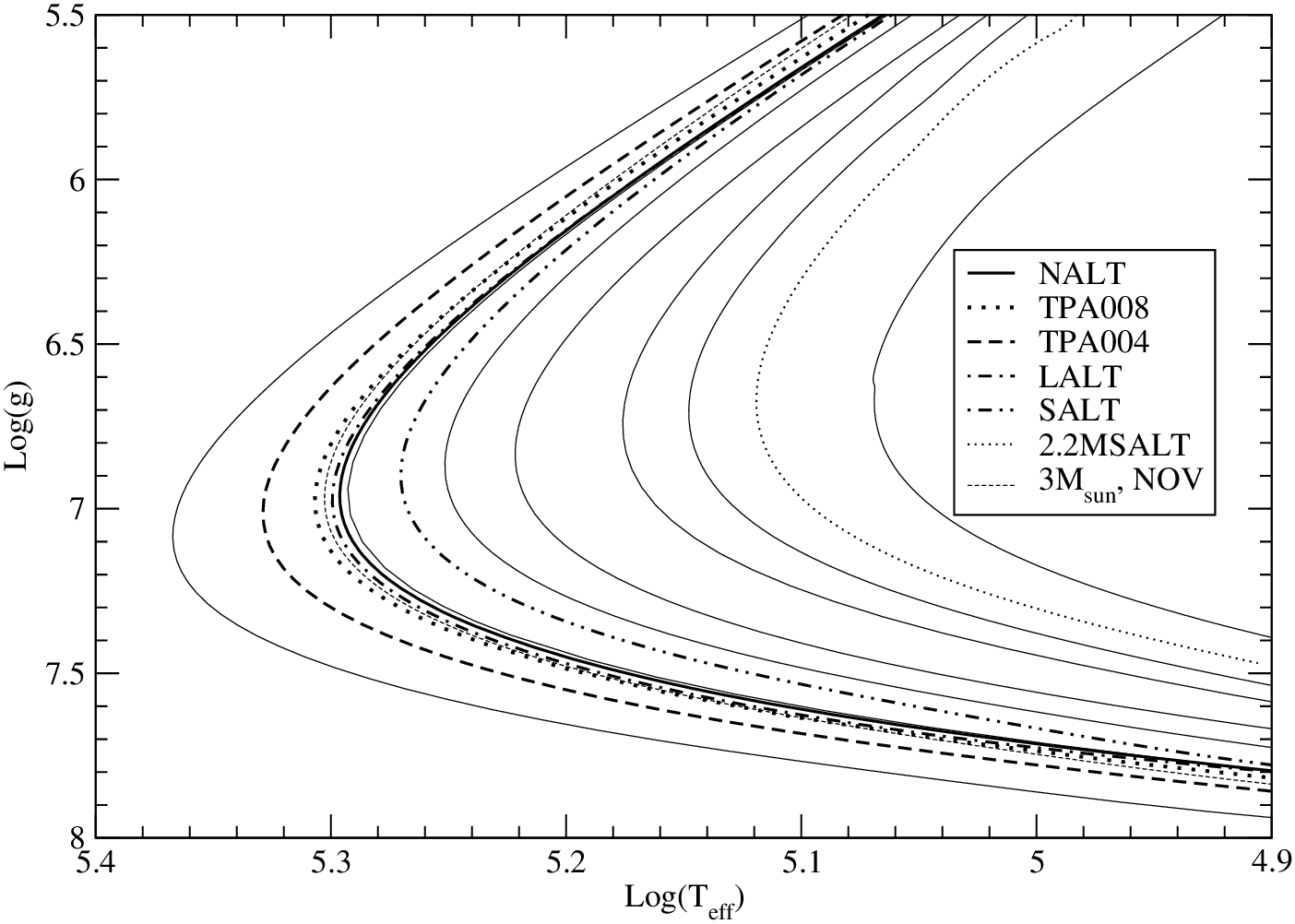} \caption{PG1159 tracks of
  this work as compared with those of MA06. Thin solid lines correspond to the
  standard ($f$=0.016 at all convective borders) tracks of MA06 with
  stellar masses of (from right to  left) 0.512, 0.53, 0.542, 0.565, 0.585,
  0.609, 0.664 \msun.}
\label{fig:gteff_OV} 
\end{figure}

As mentioned early, different mass--loss rates can also change the
initial-final mass relation of the sequences, leading to final remnants with
very different previous evolution but similar final mass. In this connection,
we have computed the evolution of an initially 2.2-\msun\ sequence by assuming
an extreme mass--loss rate during the whole AGB (sequence 2.2MSALT). As a
result, this sequence underwent only 5 thermal pulses on the AGB --- as
compared with the 15 AGB pulses of the 2.2-\msun\ sequence in MA06. The final
mass of the remnant is of 0.516 \msun, much lower than the 0.565 \msun\ quoted
in MA06.  The track for this sequence in the \gteff\ plane is shown in
Fig. \ref{fig:gteff_OV} together with the other sequences of this work and
those of MA06. Note that the 2.2MSALT track is more luminous and hotter than
that of the standard sequence of similar mass (the 0.512 \msun\ sequence in
MA06).  Note that in this case, the shift in the $M$-$g$-\teff\ relation of
the remnants would imply a
\emph{decrease} of $\sim 0.01$ \msun\ in spectroscopical masses. This value is
unexpectedly low in view of the fact that the two standard sequences in the
same region of the \gteff\ diagram have a very different previous
evolution. Indeed, the 0.512 and 0.53 \msun\ sequences in MA06 have been
calculated from an initially 1-\msun\ progenitor that went through the helium
core flash. Also, the 0.512 \msun\ has a very different intershell and surface
composition with only 2\%, by mass, of oxygen; see MA06 for a description of
this sequence. Again, it is interesting to look at the structure of the HeFC
to understand this change. As can be seen in Fig. \ref{fig:Hefreecore} (black
star symbol), although the mass and radius of this model fall almost in the
standard locus (the thick grey line in figure 2), its HeFC mass ($\sim0.45$
\msun) is significantly higher than that of the standard 0.512 \msun\ sequence
($\sim0.43$ \msun) and thus should be, again from shell homology arguments,
more luminous than the standard sequence. Indeed that is what actually
happens. Even more, the 2.2MSALT sequence has a HeFC mass that falls almost in
the middle of that of the 0.512 and 0.53 \msun\ MA06 sequences and its track
in the \gteff\ diagram does exactly the same. These considerations seem to
support the idea that is the HeFC structure --- and not the HFC --- which is
important to understand H-deficient post--AGB tracks.

So, although the findings of this section confirm that different TP-AGB
lifetimes may result in changes in the post--AGB tracks, we find that this
effect is not enough to account for the mass discrepancy mentioned in the
introduction. Indeed, we find that the PG1159 spectroscopical masses
inferred from the MA06 post-AGB tracks would be higher by at most $\sim0.015$
\msun\ (for stars close to the 0.6 \msun\ tracks) if in their calculations
MA06 had considered much shorter TP-AGB lifetimes during the
progenitor evolution of their PG1159 sequences. On the other hand we
find impossible to get a similar shift for stars close to the 0.512
\msun\ track. This is so because the lack of 3DUP in low mass stars.


\subsection{Effect of the third dredge-up efficiency}

To explore the role of the 3DUP efficiency during the TP-AGB in the
location of post--AGB tracks, we have followed the TP-AGB evolution
for three different values of the overshooting efficiency ($f$) at the
pulse driven convection zone (PDCZ) that develops during each He-shell
flash.  As shown in Herwig (2000), higher $f$ values at the bottom
of the PDCZ lead to more intense helium shell flashes and more
intense third dredge up events, while the value of $f$ at the base of
the convective envelope only plays a secondary role in determining the
3DUP efficiency (the reasons for this are extensively discussed and
shown in sections 4 and 5 of Herwig 2000).  We have, thus, calculated three
different sequences for a 3-\msun\ progenitor by adopting values of
$f$=0.016, 0.008, 0.004 at both convective borders of the PDCZ,
from now on sequences NALT, TPA008 and TPA004; sequence NALT
corresponds to that previously described. At any other convective zone
--- for example the AGB convective envelope and the core burning
regions in the previous evolution --- the ``standard'' value of
$f$=0.016 has been used. We stress that the ``standard'' value
$f=0.016$ comes from the fitting of the width of the main sequence
(Herwig et al. 1997), and thus is appropriate for the core H-burning
zone. But it may be unrealistic for the conditions at the PDCZ (Herwig
2004). All of these sequences have similar TP-AGB lifetimes. This
enables us to disentagle the 3DUP effect from the one coming from
different TP-AGB lifetimes, which was studied in the previous
section. Also, for these three sequences, mass loss during the last
interpulse phase has been artificially set in order to obtain a VLTP
and consequently a H-deficient post--AGB remnant.

\begin{figure}[ht!]
  \includegraphics[clip, width=250 pt]{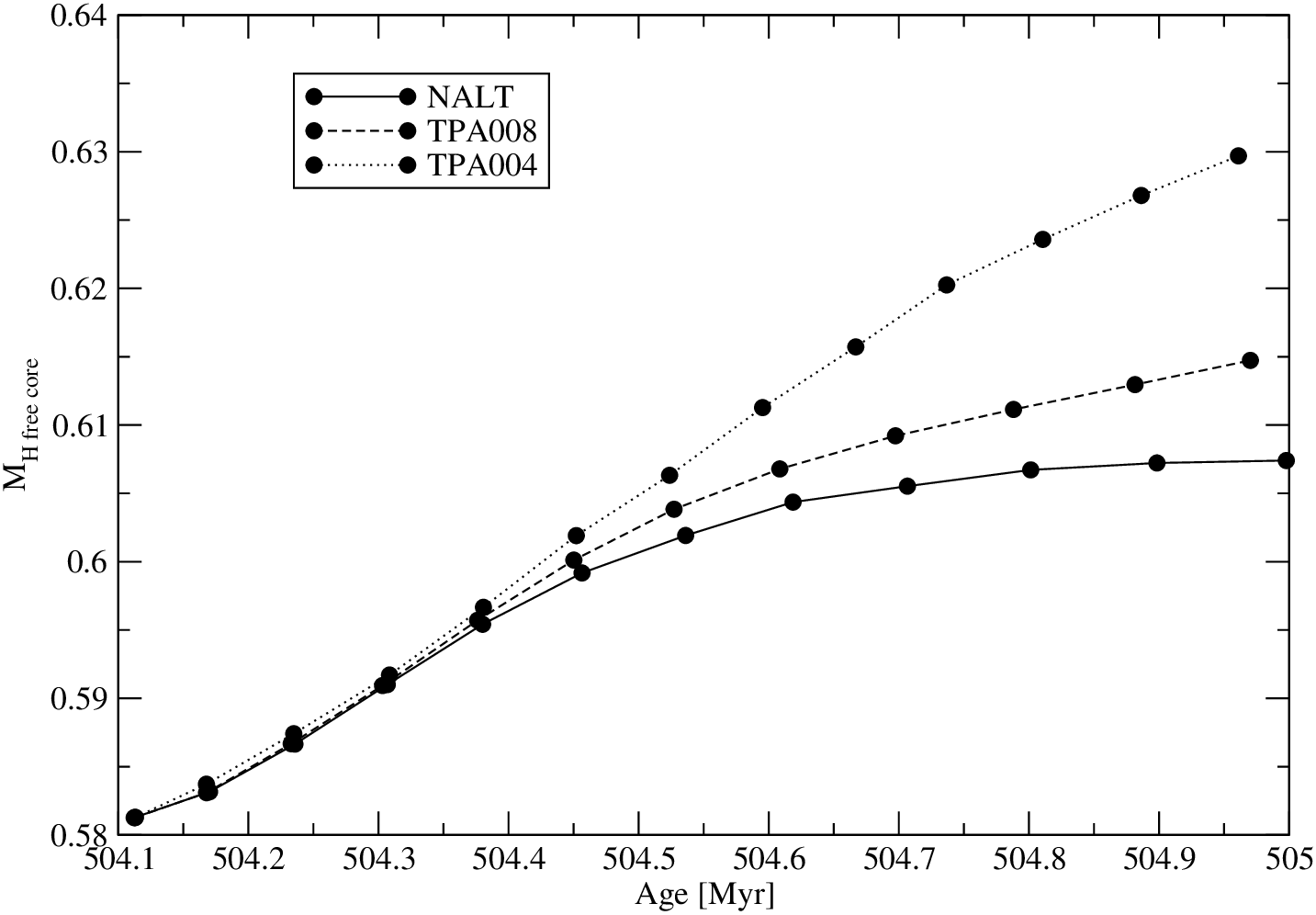}
  \caption{HFC evolution during the TP-AGB for three sequences
  with different $f$ values at the PDCZ (masses in \msun).}
\label{fig:Hfreecore-evol} 
\end{figure}

Note from Fig. \ref{fig:Hfreecore-evol} that different values of $f$
yields different evolution of the HFC. For models with higher
3DUP efficiencies the ``effective'' growth of the HFC is
stopped. This is because the increase in the HFC induced by the
H-burning shell is compensated for by a decrease during the 3DUP events. Not
only the HFC mass is altered but also, as expected, the surface and
intershell abundances --- in particular the O intershell abundance; see Herwig
(2000). As a result of the different adopted 3DUP efficiencies, the final
remnant masses are different, being 0.607, 0.617 and 0.633 \msun\ for NALT,
TPA008 and TPA004 respectively. It is worth noting that TPA004 hardly
undergoes any 3DUP events. So this sequence should be representative of
the case in which no overshooting is considered at the PDCZ.

Our results suggest that different 3DUP efficiencies do not seem to lead to an
important shift in the location of the theoretical post--AGB models in the
$M$-$g$-\teff\ space. Indeed, sequences TPA008 and TPA004 are located in 
the zone of the \gteff\ diagram corresponding to remnants of similar mass 
of the standard sequences; see  Fig. \ref{fig:gteff_OV}. 
A quantitative measure of the possible shift in the tracks relative to the 
standard MA06 ones is given Table 1. Note that there is a small
shift of 0.005 \msun\ in the derived mass for the NALT sequence as 
compared to the actual one --- we remind that NALT sequence has
the same overshooting prescription than that assumed in MA06. This is 
probably due to a combined effect of a different number of
thermal pulses and slightly different envelope composition --- which leads to
different intershell opacities, see Sect. 3. 
Because the three sequences have similar TP-AGB lifetimes, this small shift
should be taken as the level of uncertainties in these
comparisons. Keeping this in mind, the masses derived for TPA008 and 
TPA004 are practically similar to the
actual masses of these sequences. This shows that, at
least around $\sim 0.6$ \msun, the theoretical $M$-$g$-\teff\  relation
of  the MA06 H-deficient post-VLTP sequences does not seem to depend 
on the intensity of 3DUP events. This
can be understood from Fig. \ref{fig:Hefreecore}. Note that
HeFC mass-radius values of sequences TPA008 and TPA004 at the moment of the
VLTP lie on the same locus than the standard MA06 sequences of similar
masses. Thus, according to shell homology relations,
the He-shell luminosity-mass relation for these sequences should be similar 
to the MA06 ones --- which do experience efficient 3DUP events. 
Finally, we mention that the central values of density and temperature 
($T_c$, $\rho_c$) show that the HeFC
readjusts its structure to the new mass after each thermal pulse. 
At the end of the TP-AGB the $T_c$, $\rho_c$ values of TPA004 are within those
of NALT --- of final HeFC mass 0.572 \msun --- and those of the 3.5 \msun\ 
sequence of MA06 --- of final HeFC mass
0.638 \msun ---, a fact which is consistent with the final HeFC mass of
0.601 \msun\ that characterizes sequence TPA004.

\subsection{Evolution previous to the AGB}

 We explore now the effect of overshooting efficiency during both the
early AGB and the core He-burning phase on the location of the
post--AGB tracks. This bears also some relevance on the fact that, as
inferred from the two previous sections, neither the TP-AGB lifetime
nor the 3DUP efficiency are the reasons for the fact that the MA06
tracks are markedly hotter than the older H-rich tracks (Bl\"ocker
1995b).  To assess these issues, we have calculated the evolution of
an initially 3-\msun\ progenitor but \emph{without} overshooting
mixing at \emph{any} convective border of the star during its whole
evolution.  After 19 thermal pulses, a H-deficient post--VLTP sequence
of 0.626
\msun\ is obtained --- early AGB and TP-AGB lifetimes are 
$\sim4.8\times10^7$yr and $\sim2.1\times10^6$yr, respectively. This is similar
to one of the sequences of Bl\"ocker (1995b) that consisted of an initially
3-\msun\ model that after 20 thermal pulses ends its post--AGB evolution as a
0.625 \msun\ remnant --- with early AGB and TP-AGB lifetimes of
$\sim7.\times10^7$yr and $\sim1.9\times10^6$yr, respectively --- and will
allow us for comparison. The main evolutionary difference between both
sequences is the occurrence of a VLTP in the post-AGB evolution of our
sequence.

The resulting H-deficient post--VLTP track is very similar to the MA06 one and
thus much hotter than the old, H-rich, Bl\"ocker's 0.626-\msun\ track. In fact
if we estimate its mean mass from comparison with the standard MA06 sequences
we get almost the actual mass (see Table 1). The mayor difference is that the
model is slightly cooler at the knee --- a shift that would affect
spectroscopical masses less than $\sim 0.01$ \msun. From
Fig. \ref{fig:Hefreecore} we can see that the evolution of the HeFC
mass-radius relation is different from that of the standard sequences. But
even in this case the difference in the radius of the He-free core at the
moment of the VLTP amounts to only a 4$\%$ as compared with the standard
sequences of similar mass.  Consequently, it should not be surprising that the
tracks are similar.

This shows that the $M$-$g$-\teff\  relation for the post--AGB tracks
is not significantly affected by the previous evolution. Thus,
differences in the previous evolution do not seem to provide a possible 
solution to discrepancy between asteroseismological and spectroscopical 
masses nor an explanation to the difference with older tracks.

\section{The role of microphysics and composition in the location of
  post--VLTP tracks}

We explore now the importance of microphysics
and chemical compositions. Specifically we assess the effects of
changing the equation of state (EoS), chemical composition of the C-O
core and opacities --- both radiative and conductive. Here, we \emph{
do not} calculate new evolutionary sequences from the ZAMS to the
PG1159 stage; instead we consider some post--VLTP sequences of MA06
and alter their microphysics before entering the PG1159
stage. 
We have checked that the models are already relaxed to the new
physics well before reaching the knee in the HR and \gteff\
diagrams. We check this by first doing the changes at different times
in the post--VLTP evolution. We find that the tracks do not depend on
the exact moment the changes are done, thus suggesting that the
structure has already relaxed to the new situation. Additionaly, we
estimate the thermal relaxation time of the envelope as
$\tau\sim\int^{M_\star}_{M_{\rm e.b.}} c_v T dm\, /\, L_\star$ ---
where $M_{\rm e.b.}$ stands for the mass coordinate at the bottom of
the envelope. We concentrate on the 0.53 and 0.584 \msun\ remnants of
MA06. For these sequences we find that $\tau$ is about one order of
magnitude lower than the time it takes the remnants to evolve from
\lteff $\sim 4.6$ to the knee in the HR diagram. In fact, $\tau$ is
about 2500 and 1600 yr for 0.53 and 0.584 \msun\ remnants
respectively, as compared with the $\sim23000$ yr and $\sim12000$ yr
it takes the remnant to evolve to the knee. This guarantees that the
envelope is thermally relaxed at that point.

\subsection{Equation of state and chemical composition of the C-O core}
To analyze the importance of the C-O core composition and EoS we considered
the 0.53 \msun\  post--VLTP sequence from MA06. With regard to the EoS we
compare tracks resulting from the use of the standard EoS of LPCODE  --- see
Althaus et al. (2005) for references --- with those coming from an 
updated version
of Magni \& Mazzitelli (1979) EoS. The latter is a more detailed
equation of state which takes into account  non-ideal corrections such as
the pressure effects on ionization and includes Coulomb
interactions also in the non-degenerate regime --- our standard EoS only 
includes Coulomb corrections in the degenerate regime. To analyze the role
of the core composition we reset the abundances below
ln$(1-m(r)/M_\star)=-2.4$ --- thus not altering the
composition at the He-burning shell--- to two extreme values: 92\% of C and
92\% of O by mass. This is not consistent with previous
evolution but allows us to estimate its importance for post--VLTP tracks.

Both changes in the EoS and the composition of the CO core do not yield
significant changes in the post--VLTP tracks. In fact in none of these cases
do we find the shift in \lteff\ to exceed 0.01 dex, being generally much
smaller. Consequently, neither the C-O core composition nor the EoS assumed in the
computation of post--VLTP sequences play a role in the derivation of
spectroscopical masses, and we can discard these two factors as possible
reasons for a shift in post--VLTP tracks.

\subsection{Opacities}
Because the outer structure
of PG1159 stars is completely ruled by radiative transport of energy, changes
 in the opacities could yield
differences in the tracks. This may be particularly interesting as the
sequences of MA06 have been calculated for radiative opacities with a solar
scaled metallicity and PG1159 stars are known to present surface abundances
rich in s-process elements and iron deficient (Miksa et al. 2002, WH06).
In this regard, by how much the transformation of iron
into heavier elements may alter the opacities in PG1159 is not known --- for
example, in a different context, Jeffery \& Saio (2006) find differences in the
pulsation properties of subdwarf stars depending on whether it is iron or
nickel that it is enhanced. Also the exact value of the original metallicity
of the progenitor star of PG1159 is not known. We analyze the effect of
changing both radiative and conductive opacities with very different results
in each case.  Full calculations of the VLTP and post--VLTP by means of
consistent opacities are out of the scope of this work, however we can try to
get an idea of how much the opacities affect the post--AGB tracks by
artificially changing the opacities in the post--VLTP evolution by arbitrary
factors or by adopting different opacity tables.

As a result of these experiments we find that for conductive opacities
even a change of 3 orders of magnitude do not produce significant
changes in the post--VLTP tracks. Quite on the contrary, the tracks
are more sensitive to radiative opacity changes. In fact we find that
--- for both the 0.53 and the 0.584 \msun\ sequences --- increasing
the opacities by a factor 1.5 produces a reduction of $\sim 0.04$ dex
in \teff\ and $\sim 0.2$ dex in log$L/L_\odot$. Similarly a reduction
in the opacity by a factor 0.5 leads to increases of $\sim 0.075$ dex
and $\sim 0.3$ dex in temperature and luminosity, respectively. These
are important changes and would clearly affect spectroscopic mass
determinations\footnote{ It is worth noting that we do not expect
important changes in asteroseismological inferences due to changes in
opacities. This is so because asteroseismological determinations are
usually based on adiabatic period studies, which are barely
affected by changes in the opacities.}. The shift in the location of
post--VLTP tracks due to changes in the opacities is displayed in
Table 2, where we show the change in
\lteff\ for different values of $g$ and for two different remnant
masses (0.53 and 0.607 \msun). Also the induced shift in the mass
derived from comparison with the $g$ \teff\ values of MA06 tracks is
shown. Two things deserves comments. The effect of different radiative
opacities is much larger for higher remnant masses and at larger
luminosities (i.e. lower gravities).  Indeed, note that for the 0.53
\msun\ remnant an increase in the opacity of 50\% would not produce a
shift of more than 0.01 \msun\ in spectroscopical mass determinations,
and for the 0.607 \msun\ remnant the increase in the spectroscopical
mass becomes very important, reaching up to 0.07 \msun\ at high
luminosities.  Note also that the shift in \lteff\ is almost the same
for the same change in $\kappa$ regardless of the mass.


\begin{table*}[ht!]
\begin{center}
\begin{tabular}{|c|c|c|c|c|c|}\hline
Sequence           & $g=5.5$ & $g=6$  &  $g=6.5$  & $g=7$ & $g=7.5$  \\\hline \hline
0.53 \msun         & 0.0782  & 0.0764  &  0.0754   & 0.0735 & 0.0563   \\
($\kappa\times0.5$)& (-0.1072)& (-0.0687)& (-0.045)   &(-0.0292)&(-0.0188) \\\hline

0.53 \msun         & -0.0462  & -0.045  &  -0.044   & -0.0432 & -0.0365   \\
($\kappa\times1.5$)& (0.0096) & (0.0106)& (0.0099)   &(0.0086)&(0.0066) \\\hline

0.607 \msun         & 0.0788  & 0.0732  &  0.0724   & 0.0724 & 0.0757   \\
($\kappa\times0.5$)& (-0.1748)& (-0.126)& (-0.0888)    &(-0.0545)&(-0.031) \\\hline

0.607 \msun         & -0.0455  & -0.0442  &  -0.0425   & -0.0418 & -0.0422   \\
($\kappa\times1.5$) & (0.0684)& (0.0484)& (0.0366)  & (0.0243)& (0.0158) \\\hline

\end{tabular}
\label{tab:opac}
\caption{Shifts in effective temperature ($\delta$\lteff) induced by changes in $\kappa$ for
different values of $g$. The value between brackets is the predicted induced
shift in spectroscopical masses (in \msun).}
\end{center}
\end{table*}
Due to the importance of this effect we consider interesting to
analyze if the effect is due to the value of the opacity at some
speciffic region of the star --- e.g. the He-burning shell. We proceed
then to make localised changes in the opacity and found, against
expectations, that it is not the value of the opacity (per unit mass
$\kappa$) at some particular region that is relevant but the total
opacity of the envelope ($\int_{\rm envelope} \kappa \,dm$). By
looking at the models, we find that altering the radiative opacity
produces almost no change in the structure of the envelope. Then, as
$dT/dm$ is not altered by changes in $\kappa$, varying $\kappa$ leads
to an opposite  and proportional change in the
luminosity $l(m)$ of the star via the relation
\begin{equation}
\frac{dT}{dm}=\frac{-3}{64\pi^2ac} \times \frac{\kappa l}{r^4 T^3}
\label{temp_prof}
\end{equation}
A clue of why only $l$ reacts to a change in $\kappa$ can be obtained from the
following simple analytical argument. If we assume that the envelopes of these
objects are nearly homological to each other, then we have that under homology
changes (with $x=\delta r/r$) the change in the pressure and density
of a shell is (see Kippenhahn \& Weigert 1990 for a deduction)
\begin{equation}\frac{\delta P}{P}=-4x,
\,\frac{\delta \rho}{\rho}=-3x.
\end{equation} 
Then if we assume an ideal gas equation of
state for the envelope --- which is quite correct --- we have the additional
relation 
\begin{equation}
\frac{\delta T}{T}= -\frac{\delta \rho}{\rho}+\frac{\delta P}{P}=-x.
\end{equation}
 Using the equation of the temperature profile (Eq.\ref{temp_prof})
by imposing an arbitrary change in the opacity $\delta \kappa/\kappa$
and using 
\begin{equation}
\delta\left(\frac{dT}{dm}\right)=\frac{dT}{dm} \times \frac{\delta T}{T}
\end{equation}
we get 
\begin{equation}
\frac{\delta T}{T}=\frac{\delta \kappa}{\kappa}+\frac{\delta l}{l}-4
\, \frac{\delta r}{r}-3 \, \frac{\delta T}{T}
\end{equation}
and by using Eq. 2 and 3 we finally find
\begin{equation}
\frac{\delta l}{l}= -\frac{\delta \kappa}{\kappa},
\end{equation}
 which is quite similar to what it is observed in the numerical models
--- during the horizontal part of the tracks.  Let us note that this
change in the luminosity does at first order balance (in
Eq. \ref{temp_prof}) the change in opacity, leaving only second order
effects on the factor $\kappa l$ to be balanced by the other factors
in Eq. \ref{temp_prof}:
\begin{equation}
\kappa^,l^,= \kappa l \left(1+\frac{\delta
  \kappa}{\kappa}\right)\left(1+\frac{\delta l}{l}\right)=\kappa l \left(1-\left(\frac{\delta
  \kappa}{\kappa}\right)^2\right).
\end{equation}
Also, note that due to the high powers of $r$ and $T$  in
Eq. \ref{temp_prof}, small changes in these quantities should be enough to
balance the remaining second order effects.  In fact, when looking at the
numerical models all $r(m), P(m), T(m)$ remain almost unchanged by the change
in the opacity, being $l(m)$ the only structure variable that undergoes an
important variation. The change in $l(m)$ seems to be associated with 
a change in
the energy liberated by the helium burning shell ---change which can be
attained with almost no change in $T(m)$ due to the extremely high sensitivity
of triple alpha reaction rates with temperature--- and by a change in the
$dS/dt$ term of the energy equation --- the lower the opacity, the faster the
evolution and contraction of the envelope. Sumarizing we can say that
altering the radiative opacity of the envelope leads to a similar change in
the $l(m)$ profile of the star which balances (at first order) the effects of
the opacity change in Eq. 1. It seems that as consequence of this balance
only minor changes appear in the other structure variables which remain 
almost unchanged --- this probably reflects the fact that the
run of these variables in the envelope of the star is forced by the radius and
mass of the He-free core where most of the gravitational field is
generated. Although this does not intend to be a complete explanation,
something which is impossible nowadays due to the lack of an accepted
explanation for the behaviour of structures with burning shells \footnote{In
fact the problem is in some way related with the long standing problem of why
stars become red giants; see Faulkner (2005) for a recent discussion, and
review, of this issue.}, we think that it sheds some light on what is
happening on these models. Finally let us mention that as $r(m)$ is not
changed, then the radius of the star is almost the same, independently of the
value of the opacity. Then, due to Steffan-Boltzmann's law we have that the
change in the opacity produces a variation in the effective temperature of
\begin{equation}
\frac{\delta T_{\rm eff}}{T_{\rm eff}}=\frac{1}{4}\frac{\delta
  L_\star}{L_\star}- \frac{1}{2}\frac{\delta R_\star}{R_\star}\sim
  \frac{1}{4}\frac{\delta L_\star}{L_\star}
\end{equation}
In fact for the 0.584 \msun\ models with normal and enhanced (for a
factor 1.5, $\frac{\delta\kappa}{\kappa}\sim0.5$) opacity we have
that, in the knee, $\frac{\delta L}{L}\sim 0.34$ and $\frac{\delta
T_{\rm eff}}{T_{\rm eff}}\sim 0.09$ and thus $\frac{\delta T_{\rm
eff}}{T_{\rm eff}}\sim
\frac{1}{4}\frac{\delta L_\star}{L_\star}$. Note however that $\frac{\delta
L}{L}\sim 0.34$ is far from the value 0.5 in
$\frac{\delta\kappa}{\kappa}$ ($\frac{\delta L}{L}$ is close to 0.5 in
the horizontal part of the track, i.e. \lteff$< 4.8$).

\subsection{Effect of different compositions on the opacities}
As models seem to be sensitive to the value of the radiative opacity
in the envelope, we have analyzed how much opacities can change due to
different adopted compositions. 

Firstly, we have assessed possible changes
in the tracks due to changes in the total amount of metals in the
models --- with the exception of C and O which are always kept consistent
with the envelope abundance. We did this by using OPAL C- O- enhanced 
tables for $Z=0.01$ and $Z=0.03$ --- all of our previous
sequences correspond to $Z=0.02$. The change in metallicity was done 
well before  evolution reached the
PG1159 stage. 
 We find that the resulting shifts in the tracks are barely noticeable.
In fact, at the knee in the HR diagram the $Z=0.01$ and $Z=0.03$ tracks
differ in effective temperature by only  $\sim 0.006$ dex.

\begin{figure}[t]
  \includegraphics[clip, width=250 pt]{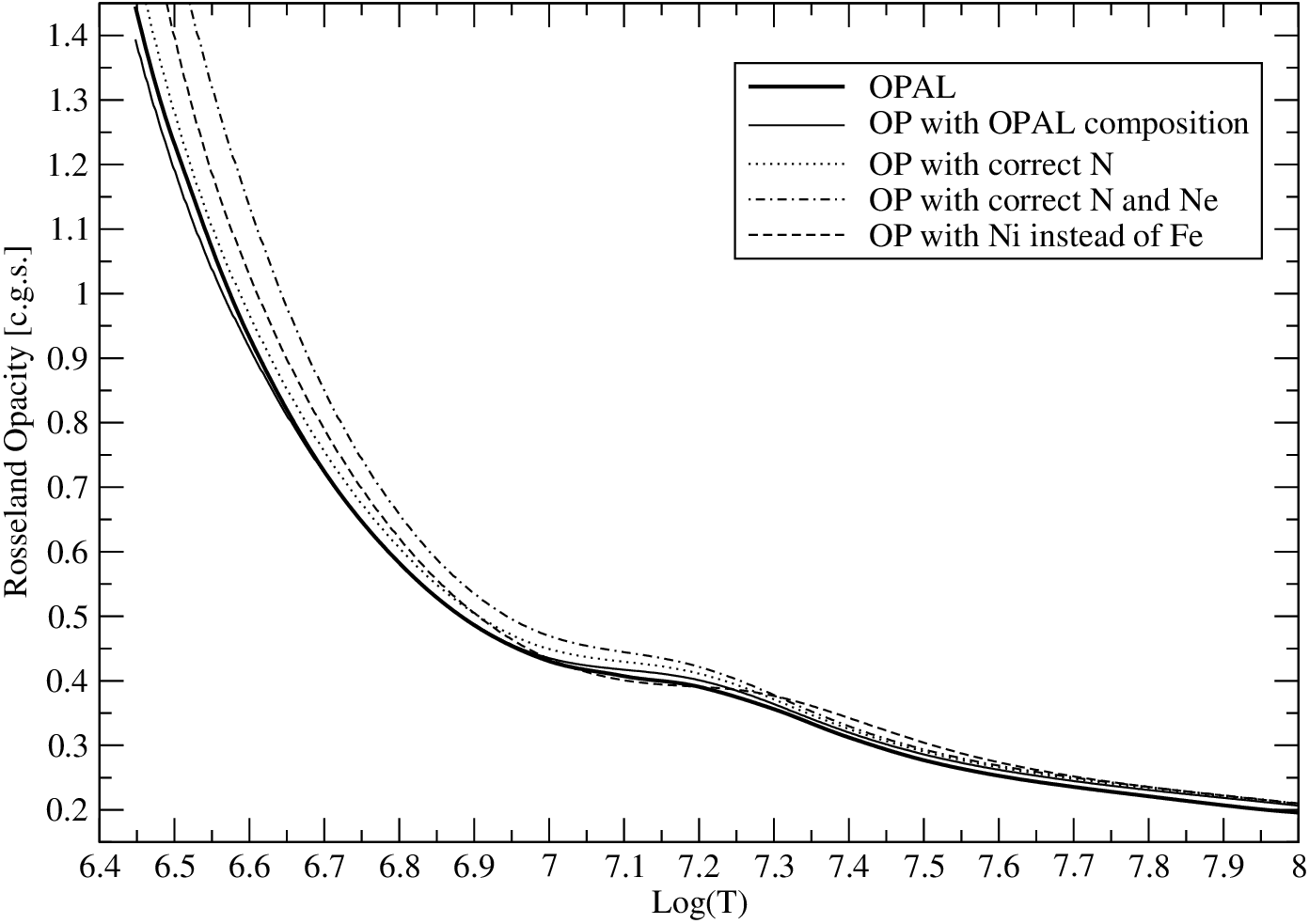}
  \caption{Value of the opacity for different adopted compositions. }
\label{fig:OP-OPAL} 
\end{figure}

Secondly, we have explored the use of opacities fully
consistent with the abundances of the models. This
is not a minor issue as Ne and N can be much larger than their solar
scaled values --- also Mg can reach values of 2\%; see Werner et
al. 2004. Specifically, we have used the tool at OP project website
(Badnell et al. 2005) which allows to calculate opacities for
arbitrary compositions. In this case we have not made any track
calculation but instead we have just compared the opacities for a given
model (i.e. for a given $T$ and $\rho$ profile). We compare first OP and
OPAL opacities for the same imput composition. The result is shown in 
Fig. \ref{fig:OP-OPAL}
Note that, for  log $T>7$,  OP opacities tend to be about 5
to 7 \% larger. This would probably introduce a small change of
about $ 0.01$ dex in the tracks. The
inclusion of N and Ne --- with abundances consistent with those displayed by
the stellar models ---  markedly
increases the opacity values below log $T\sim7.3$, but
almost no changes are present at higher temperatures where most of
the mass of the envelope is stored, see Fig.\ref{fig:OP-OPAL}. 
As PG1159 stars are supposed to be iron deficient due to s-process (WH06), we
have analyzed the extreme case for which all the iron was changed into
Ni. In this case, the opacity bump is located at larger $T$ values, thus
increasing  opacity  between
log $T$= 7.3 and 7.6. Although this change in the opacity is not enough
to reconcile the discrepancy between spectroscopical and asteroseimic 
masses of PG1159 stars,
it is important to note  that modifying the heavy metal distribution does
introduce a change in the opacities at high temperatures. Indeed, as
opacities increase with the atomic number of the elements --- due to the
increase in the possible atomic transitions, Roger \& Iglesias (1994) --- it
remains to be seen to what a degree an important increase in the content of
very heavy metals due to s-process (both in the AGB and at the VLTP)
can increase the opacity at the bottom of the envelope. Note that
because of the higher ionization potentials, those elements are expected to
affect opacities at much higher temperatures than do Fe or Ni.

\section{Other issues}
\subsection{External consistency}
We have compared our tracks with the few models of modern H-deficient post--AGB
sequences available in the literature, particularly the 0.604 \msun\  model 
of Herwig (2005) and the 0.632 \msun\  model of Lawlor \& Mac Donald (2006).
\begin{figure}[ht!]
  \includegraphics[clip, width=260 pt]{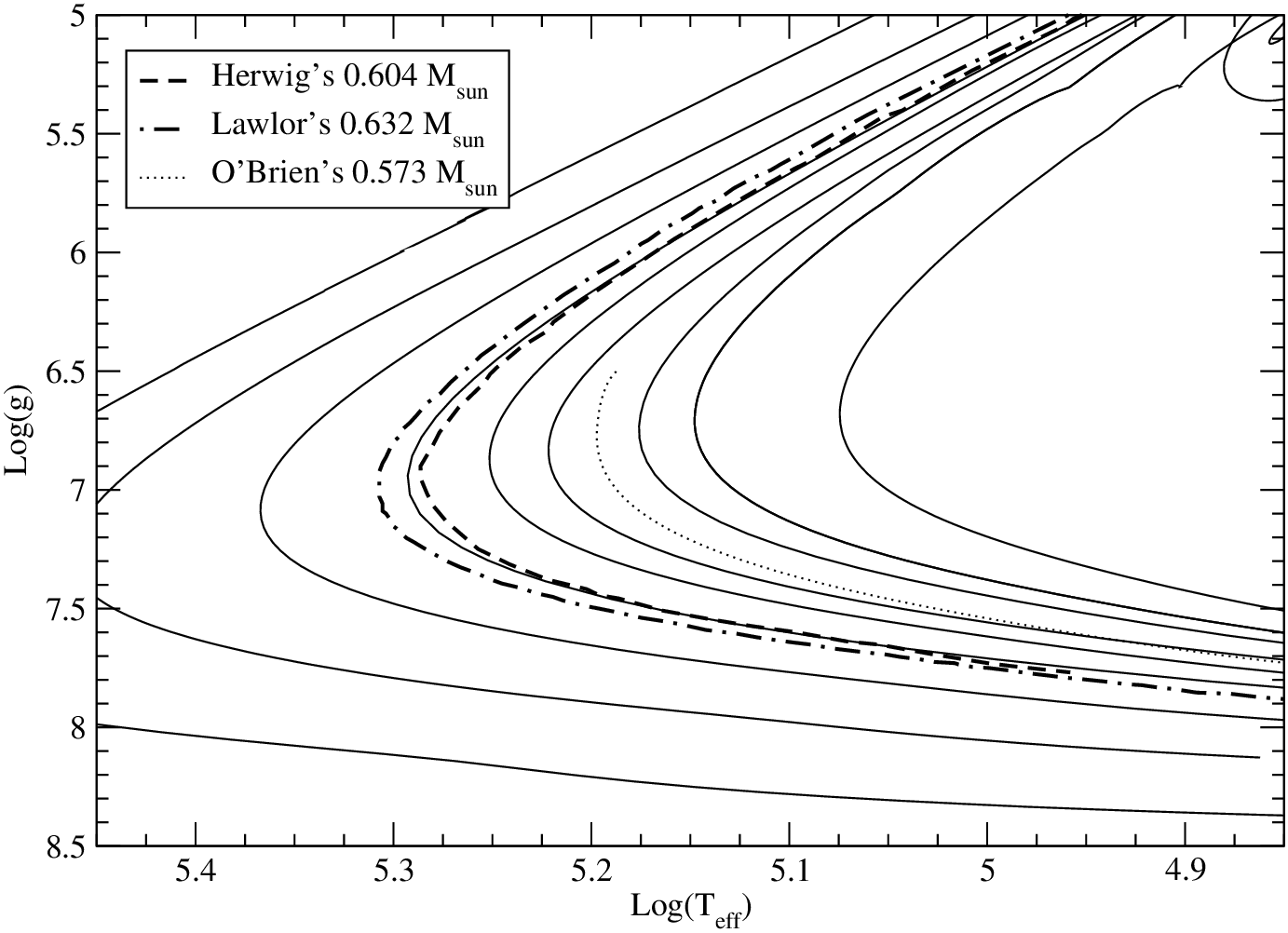}
\caption{H-deficient tracks of Herwig (2005) and Lawlor \& Mac Donald
  (2006) compared with our post--VLTP tracks (with masses 0.87, 0.741, 0.664,
  0.609, 0.585, 0.565, 0.542, 0.530, and 0.515 \msun). Also a non--late helium
  flash (but H--deficient) track from O'Brien (2000) is shown for comparison. }
\label{Herwig-Lawlor} 
\end{figure}

Note  from Fig. \ref{Herwig-Lawlor} that our models show a good agreement
with the tracks of both authors. The agreement is remarkable despite
the very different input physics and evolutionary history of progenitor
stars considered by those authors. Indeed,  Lawlor \& Mac
Donald (2006) models do not include any kind of overshooting prescription and
Herwig's model is the result of an initially 2 \msun\ star model and thus 
with a distinct previous evolution than our 0.609 \msun\ sequence 
which comes from a
3 \msun\ model. In addition, the EoS are different in all the cases. 
This supports the findings in Sects 2 and 3. 
For a quantitative inference, we estimate
masses for those tracks by comparing their relative location with MA06
tracks. We derive masses of about 0.611 \msun\ and 0.623 \msun\ for Herwig and
Lawlor \& Mac Donald sequences respectively --- note that the  resulting 
Herwig's track becomes bluer than ours, leading to slightly lower 
spectroscopical masses. In both cases the induced shift in spectroscopical 
masses would be lower than 0.01 \msun, thus reinforcing the robustness of
the MA06 post--AGB tracks.
\subsection{Comparison with H-burning tracks}
\begin{figure}[ht!]
  \includegraphics[clip, width=250 pt]{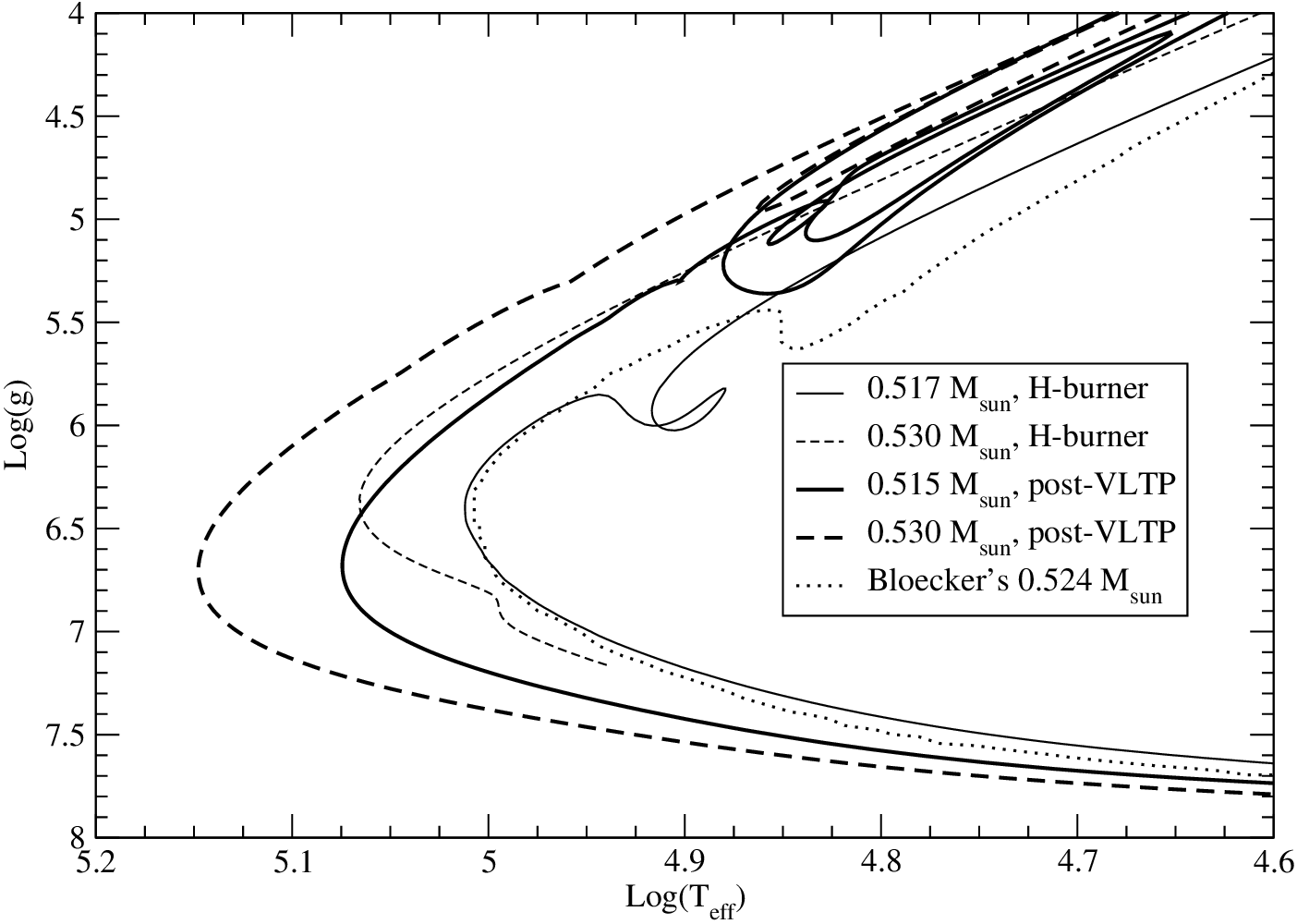}
\caption{Comparison between H-rich, H-burning tracks and H-deficient,
He-burning tracks of low mass. It is clear from the figure that H-burners have
lower \teff and are much similar to Bl\"ocker's tracks. Using H-rich tracks to
estimate PG1159 masses can lead to important overestimations in the low mass
range.}
\label{Bloecker H} 
\end{figure}

As mentioned early in this work H-burners old tracks are cooler than MA06
tracks, thus leading to a much better agreement with asteroseismology (WH06).
 As we discussed in Sect 2, the difference in the location between old and
new tracks cannot be tracked back to a distinct previous evolution, with the
exception of the VLTP event.
It is worth noting that old models (Bl\"ocker 1995a,b) are based on
the Cox \& Stewart (1970) opacities, in contrast to new models that
use OPAL and molecular opacities.  In this connection we feel
important to recall that already Dreizler \& Heber (1998) noted a shift of
0.03 \msun\ between old Wood \& Faulkner (1986) and O'Brien \& Kawaler
(as they appear in Dreizler \& Heber 1998, here shown in Fig. 6) 
helium--burning tracks,
where the latter make use of modern OPAL opacities.  Appart from
possible changes that could arise from the different 
opacities used in the calculations, the difference in the tracks can be 
expected from  the very fact that the Bl\"ocker's tracks are H-burners 
while VLTP are helium burners. In fact, we note from our sequences 
that there is a noticeable difference in H--burning-post--AGB and post--VLTP 
tracks for low remnant masses ($\lesssim 0.53$ \msun). This can be seen in 
Fig \ref{Bloecker H}, where we show our post--AGB H--burning 
tracks (with H--rich surface compositions) of 0.517
and 0.53 \msun\  compared with the post--VLTP tracks of similar mass. Note
that post--VLTP tracks are certainly bluer than their H--rich
counterparts and that they are more compact in the high gravity region
of the diagram (log $g>6.5$), as found by MA06. Also note from the same
figure that our 0.517 \msun\  track is very similar to Bl\"ocker's 0.524 \msun\  track, strongly suggesting that the difference with Bl\"ocker
(1995b) tracks at (very) low masses is because MA06 tracks are post--VLTP
tracks.  Using H-rich tracks to determine spectroscopical masses for
PG1159 stars will certainly influence the result, in particular the stellar 
masses for  high--gravity  PG1159 stars would be largely overestimated.

\subsection{Comparison of post-LTP and post-VLTP tracks}
As mentioned early in this work, $^{14}$N-deficient PG1159 objects are
probably the descendents of LTP events. Thus, a priori one should be
careful about using post-VLTP tracks for all PG1159 stars. In this
context we now turn to analyze the question if there are systematic
differences between post-VLTP and post-LTP tracks. From figure 1 of
Herwig 2001 it seems that there is no differece once the star enters
the PG1159 stage. However we will now analyze a wider range of
masses. In Fig. 8 PG1159-tracks coming from VLTP and LTP are compared
for similar remnant masses. In the upper panel LTP tracks with
different H-abundances are compared with VLTP tracks of similar
mass. The $\sim0.515$ \msun\ tracks correspond to the sequence
analysed in Althaus et al. (2007). In these sequences two different
post-LTP evolutions have been considered. The first in which the final
surface H-abundance is normal\footnote{In this sequence, due to the
very low remnant mass, the low intensity of the He-flash does not lead
to any 3DUP.} and a second in which due to mass loss episodes the
whole H-rich envelope was eroded, exposing the He-rich intershell. Due
to the absence of the H-burning shell in the second case it lies very
close to the postVLTP tracks. The second experiment is also shown in
the upper panel of Fig. 8 corresponds to an LTP sequence (0.543 \msun)
in which the total H-content of the star was artificially diluted to
different depths, thus leading to different final surface
H-abundances. As can be seen once the star reaches the PG1159 stage,
the lower the surface H-abundance the closer the LTP-tracks gets to
the VLTP track of similar mass. Finally, in the lower panel of Fig. 8
post-LTP tracks of H abundances close to the usual detection limit are
compared with VLTP tracks of similar mass. From that plot is clear
that for surface gravities above log $g=6$, where almost all PG1159
stars lie, VLTP tracks and LTP tracks with low H-abundances are
similar. Then using post-VLTP tracks for spectroscopic mass
determinations of LTP objects with no detectable H does not seem to
introduce any systematic effect on the mass determination. On the
contrary using post-VLTP tracks for hybrid PG1159 stars may produce an
important underestimation of the mass.
\begin{figure}[ht!]
  \includegraphics[clip, width=250 pt]{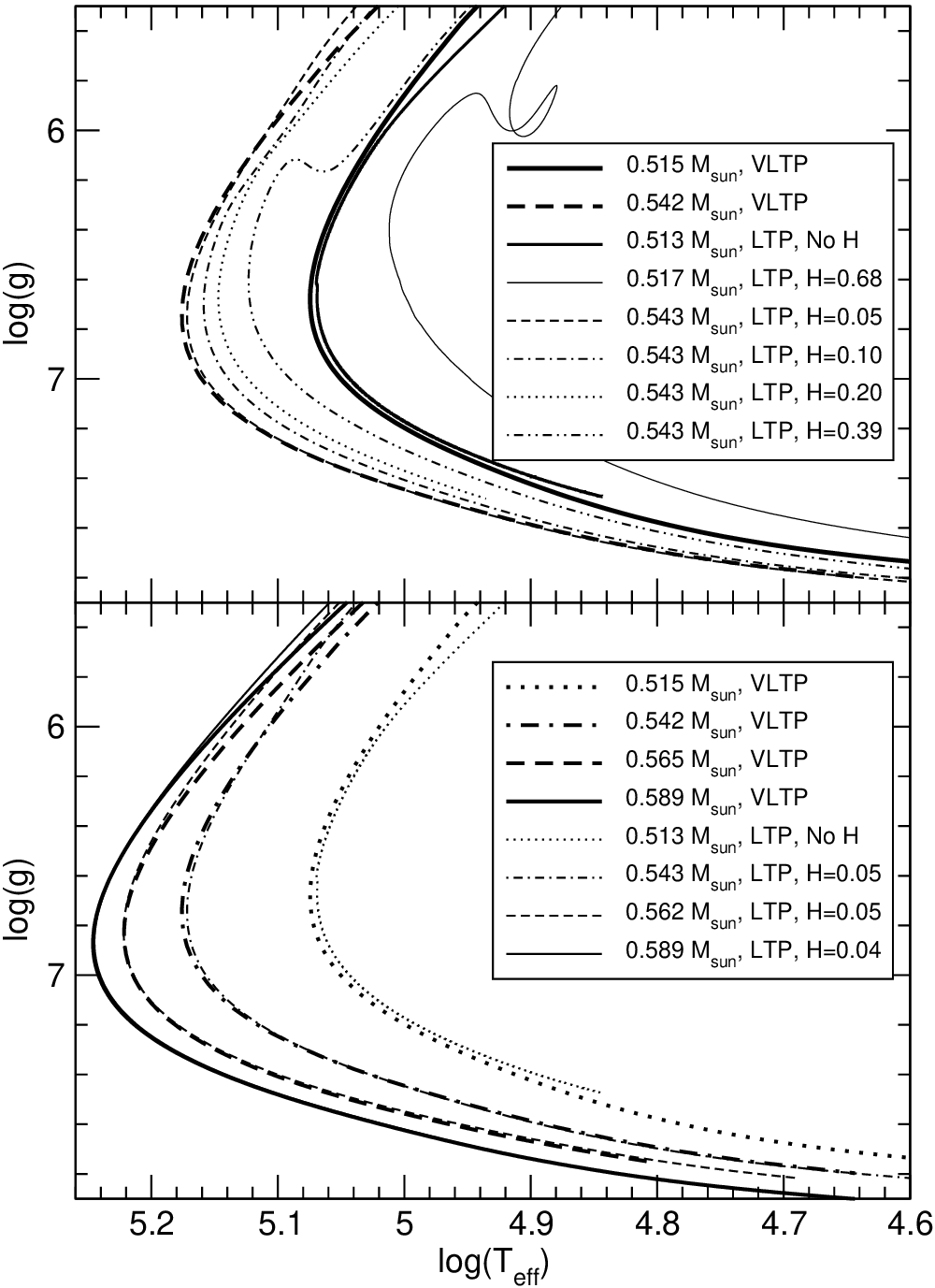}
\caption{Comparison between post-LTP and post-VLTP tracks. Upper panel: Comparison between VLTP tracks and post-LTP tracks of similar mass but different H abundances. Lower panel: Comparison between post-VLTP tracks with post-LTP tracks that display surface hydrogen abundances, close to the detection limit.}
\label{Fig:LTP-VLTP} 
\end{figure}
\section{Discussion and final remarks}
In the present work we have analyzed how uncertainties in the modeling
of H-deficient post-VLTP remnants could affect spectroscopic mass
determinations of PG1159 type stars. In Sect. 2, inspired by a
suggestion in WH06 we have analyzed the importance of previous
evolution. As the calculation of each full sequence is extremely time
consuming (both computational and human, as at some stages the models
need hand interaction to converge them) we had to restrict ourselves
to a limited region of the parameter space. Even in this case some
conclusions can be drawn. Third dredge up alone does not seem to
change the theoretical \lteff-log$g$-$M$ locus and consequently its
uncertainties can not affect spectroscopic mass determinations. On the
other hand differences in mass loss rates (i.e. TP-AGB lifetimes)
alter the location of the tracks, but only slightly. In fact our
simulations show that even a reduction by a factor of 3 of TP-AGB
lifetimes would not increase spectroscopic mass determinations by more
than $\sim0.015$\msun. Going even further we have shown that even
extreme mass losses that produce low mass remnants ($\sim0.515$ \msun
) from very different progenitors than those in MA06, does not
introduce important shifts in spectroscopic mass determinations, being
only $\sim0.01$ \msun. In an even more extreme  limiting case we
have computed a sequence, in which no overshooting was considered in
the whole evolution, and found a very similar post-VLTP track than
MA06. All these experiments suggest, contrary to the argument in WH06,
that previous evolution only plays a secondary (and not very
important) role in determining the theoretical \lteff-log$g$-$M$
locus. In light of these results we have discussed why the argument
presented in WH06 does not apply to post-VLTP sequences. We showed
that it is the HeFC (if any), and not the HFC, mass and radius that is
important for post-VLTP tracks. In particular we find that the HeFC
converges faster than the HFC in the mass-radius diagram.
However, shell homology relations (as those used to derive the
luminosity-mass-radius relation) should not be taken too seriously in
these models, as they neglect the importance of the envelope and only
relate the luminosity of the burning shells to the properties of
matter in the burning shells and to the values of mass and radius of
the core, since we find (in Sect. 3.2) an important dependence of the
shell luminosity with the whole opacity of the envelope.

We have roughly addressed the robustness of the tracks regarding EoS,
C-O core composition, conductive and radiative opacities.  We find
that only radiative opacity may affect the location of the tracks to
some an extent. Specifycally we find that the luminosity of the
post--VLTP sequences in the horizontal part of the HR diagram is very
sensitive to the envelope opacity. In fact the luminosity of the
He-burning shell turns to be sensitive to the total opacity of the
envelope.  We also present some analytical arguments to explain the
shift induced by changes in radiative opacitites.
In this connection we explore how important the envelope composition
can be for the opacity of the envelope.  We find that changes in light
metals (Ne and Mg) can make important changes in the opacities but
only at low temperatures ($T<5\times10^7$K) where no much mass of the
envelope is stored.  Although this may be important for pulsational
studies of PG1159 stars, it will certainly not change the
\teff\ of the sequences. By contrast changing Fe into Ni in the opacity
calculations we find a more slightly important change. This
particularly leaves open the question of how much opacities at the
bottom of the envelope can change if important amounts of Fe are
transformed into very heavy metals by s-processes. We can conclude
that, unless there are important changes in the abundances of very
heavy elements due to s-process, an increase in the opacity at high
$T$ is not expected to change more than 10\%.

All these arguments show that MA06 tracks are robust enough as to be
used for spectroscopical mass determinations  of PG1159-type stars
(specially at high gravities; log $g\gtrsim 6$). This robustness is
reinforced by the good agreement  (which corresponds to
differences of $\lesssim 0.01$\msun\ in spectroscopic mass
determinations) between those tracks with the other modern post--VLTP
tracks available in the literature (Herwig 2005 and Lawlor \& Mac
Donald 2006). We have also addressed in Sect. 4.3 if any systematic in
the mass determination may be due to the fact of some PG159 stars
being post-LTP objects with H-abundances below the detection limit. We
find that the resulting tracks in the PG1159 region of the \teff-g
(log $g>6$) diagram are very similar to post-VLTP tracks when surface H
abundance is below $\sim5$\% by mass fraction. Thus, we conclude that
the post-VLTP tracks of MA06 are solid enough for spectroscopic mass
determinations of post-LTP objects with H-abundances below the
detection limit and, thus, it seems that no systematic should be
present due to this effect. On the contrary, we find that using
post-VLTP tracks for PG1159 stars with important H-abundances (the so
called hybrid PG1159 stars) may lead to an important underestimation
of the mass. Regarding the difference with Bl\"ocker's H-rich
post--AGB tracks we can say that, for low mass remnants
($\lesssim0.53$\msun), the differences in the tracks seem to be mainly
due to the fact that those tracks are H-burners since our own H-burner
sequences are much colder than our post--VLTP ones. Other differences
with older tracks may be due to the difference in the opacities
adopted for the He, C -rich intershell (note that older works make use
of old Cox \& Stewart opacities). These seems to be supported by the
good agreement between all H-deficient tracks that include modern OPAL
opacities ---Herwig 2005, Lawlor \& Mac Donald 2006 and, more roughly
($\sim0.02$\msun), even with the non- late helium flash 0.573 \msun\
sequence of O'Brien 2000.

 From the present work we judge that the  systematic discrepancy
between asteroseismological and spectroscopical mass determination
methods should not be attributed to uncertainties in post-AGB
tracks. Whether the discrepancy comes from errors in
asteroseismological or spectroscopical determinations is not known,
however some points are worth emphasising. Although asteroseismology
is usually accepted as a more accurate method (very low error bars are
usually given), its robustness is not so clear. In fact recent works
(C\'orsico \& Althaus 2006, C\'orsico et al. 2007a and C\'orsico et
al. 2007b) show the results of asteroseismology to be method
dependent. In this context it is worth emphasising that the
asteroseismic mass of \object{PG 1159-035} is reduced to $\sim
0.56$\msun\ ---only $\sim0.02$ \msun\ higher than its spectroscopical
mass--- when detailed evolutionary models and averanged period spacing
(instead of the usually adopted asymptotic period spacing) are used in
the analysis, see C\'orsico et al. (2006). Interestingly enough,
during the referee stage of this article a new study of
\object{PG 0122+200} (C\'orsico et al. 2007b) which is based on our
evolutionary models and a detailed period by period fitting procedure,
reduces the mass discrepancy (with MA06 value) in this star to less
than a  4\%. This clearly shows the existence of serious systematics in
standard (i.e. based on asymptototic period spacing)
asteroseismological determinations. In this context is worth noting
that a mean PG1159 mass of 0.573 \msun\ like the one deduced from MA06
tracks, even if sensitively lower (0.044 \msun) than previously
thought, is in good agreement with that of their probable descendants,
the DB white dwarfs (0.585 \msun, Beauchamp et al. 1996)\footnote{This
is just a rough comparison as neither all PG1159 stars are expected to
evolve into DB white dwarfs nor all DB white dwarfs are expected to be
the result of single stellar evolution; e.g. see Saio \& Jeffery
(2002).}. Then our results not only call for a revision of PG1159
model atmospheres but, specially, for a revision of systematics in
usually adopted asteroseismological mass determination methods.

Our full set of evolutionary tracks for post-VLTP objects is available
at our web site at {\tt http://www.fcaglp.unlp.edu.ar/evolgroup/}.

\begin{acknowledgements}
M3B wants to thank Achim Weiss, Agis Kitsikis and Alejandro C\'orsico
for useful and instructive discussions and the Max Planck Institut
f\"ur Astrophysik in Garching and the European Assossiation for
Research in Astronomy for and EARA-EST fellowship during which part of
this work was done. This research was partially supported by the PIP
6521 grant from CONICET.
\end{acknowledgements}


\begin{thebibliography}{}

\bibitem{AS05} Althaus, L. G., Serenelli, A. M.,  Panei, J. A., et al. 2005, 
A\&A, 435, 631
\bibitem{AMC07} Althaus, L. G., C\'orsico, A. H., Miller Bertolami, M. M. 2007, A\&A, 467, 1175
\bibitem{Baetal05} Badnell, N. R., Bautista, M. A., Butler, K., et al. 2005, MNRAS, 360, 458
\bibitem{Betal96} Beauchamp, A., Wesemael, F., Bergeron, P., Liebert, J., Saffer, R. A. 1996, ASP Conference Series, 96, 295
\bibitem{B95a} Bl{\" o}cker, T. 1995a, {A\&A}, 297,727
\bibitem{B95b} Bl{\" o}cker, T. 1995b, {A\&A}, 299,755
\bibitem{CA06} C\'orsico, A. H. \& Althaus, L. G. 2006, A\&A, 454, 863
\bibitem{CAM06} C\'orsico, A. H., Althaus, L. G., Miller Bertolami,
  M. M. 2006, A\&A, 458, 259
\bibitem{CAMW07} C\'orsico, A. H., Althaus, L. G., Miller Bertolami,
  M. M., \& Werner, K.  2007a, A\&A, 461, 1095
\bibitem{CAM07} C\'orsico, A. H., Miller Bertolami,
  M. M., Althaus, L. G., Werner, K., \& Vauclair, G. 2007b, A\&A, to
  be submitted
\bibitem{CS70} Cox, A. N., Stewart, J. N., 1970, ApJS, 19, 243
\bibitem{dM02} De Marco, O. 2002, ApSS, 279, 157
\bibitem{DH98} Dreizler, S. \& Heber, U. 1998, {A\&A}, 334,  {618}
\bibitem{Eetal06} Eisenstein, D. J., Liebert, J., Harris, H., et al. 2006, ApJS, 167, 40
\bibitem{F05} Faulkner, J. 2005, ``The scientific legacy of Fred Hoyle'',
  edited by Douglas Gough, ISBN 0-521-82448-6, p. 149 
\bibitem{H00} Herwig, F. 2000, A\&A, 360, 952
\bibitem{H01} Herwig, F. 2001, {ApSS}, 275,  {15}
\bibitem{H04} Herwig, F. 2004, {ApJS}, 155,  {651}
\bibitem{H05} Herwig, F. 2005, ARA\&A, 43, 435
\bibitem{HBSE97} Herwig, F., Bl{\"o}cker, T., Sch{\"o}nberner, D., El Eid, M. 1997, {A\&A}, 324, {L81}
\bibitem{HSB98} Herwig, F., Sch\"onberner, D., Bl\"ocker, T. 1998, {A\&A}, 340, {L43}
\bibitem{HBLD99} Herwig, F., Bl{\"o}cker, T., Langer, N., \& Driebe, T. 1999, {A\&A}, 349, {L5}
\bibitem{HFW06} Herwig, F., Freytag, B., Werner, K. 2006, IAU Symp 234, 103
\bibitem{Ietal83} Iben, I., Kaler, J. B., Truran, J. W. \& Rensini, A. 1983,
  ApJ, 264, 605
\bibitem{JS00} Jeffery, C. S. \& Saio, H. 2006, MNRAS, 372, L48
\bibitem{KB94} Kawaler, S. D. \& Bradley, P. A. 1994, ApJ, 427, 415
\bibitem{KW07} Kitsikis, A. \& Weiss, A. 2007, ASP Conference Series, in press
\bibitem{KW90} Kippenhahn, R., Weigert, A. 1990, Stellar Structure and Evolution, Springer-Verlag
\bibitem{LM06} Lawlor, T. M. \& Mac Donald, J. 2006, MNRAS, 371, 263
\bibitem{MM79} Magni G. \& Mazzitelli, I. 1979, 72, 134
\bibitem{Metal02} Miksa, S., Deetjen, J. L., Dreizler, S., Kruk, J. W., 
Rauch, T., \& Werner, K. 2002, A\&A, 389, 953 
\bibitem{MA06} Miller Bertolami, M. M., Althaus, L. G. 2006,
A\&A, 454, 845 [MA06]
\bibitem{O00} O'Brien, M. S. 2000, ApJ, 532, 1078
\bibitem{RI94} Rogers, F. J. \& Iglesias, C. A. 1994, Science, 263, 50
\bibitem{JS06} Saio, H. \& Jeffery, C. S. 2002, MNRAS, 333, 121
\bibitem{W01} Werner, K.,\& Herwig, F. 2006, PASP, 118, 183 [WH06]
\bibitem{WHH91} Werner, K., Heber, U., Hunger, K. 1991, {A\&A}, 244, 437
\bibitem{Wetal04} Werner, K., Rauch, T., Barstow, M. A., Kruk, J. W. 2004, A\&A, 421, 1169
\bibitem{WF86} Wood, P. R., \& Faulkner, D. J. 1986, {ApJ}, 307, {659}

\end{thebibliography}
\end{document}